\documentclass[12pt]{article}

\newcommand{\anth}[1]{

\vspace{7pt} \noindent {\bf \underline{#1}} \hspace{3pt} }

\usepackage{graphicx}

\usepackage{amsmath}

\begin{document} 
\begin{titlepage}
\begin{flushright}

\end{flushright}

\begin{center}
{\Large\bf $ $ \\ $ $ \\
Nonlocal Charges for Bonus Yangian Symmetries of Super-Yang-Mills
}\\
\bigskip\bigskip\bigskip
{Nathan Berkovits and Andrei Mikhailov\footnote{On leave from 
Institute for Theoretical and 
Experimental Physics, 
117259, Bol. Cheremushkinskaya, 25, 
Moscow, Russia}}
\\
\bigskip\bigskip
{\it Instituto de F\'{i}sica Te\'orica, Universidade Estadual Paulista\\
R. Dr. Bento Teobaldo Ferraz 271, 
Bloco II -- Barra Funda\\
CEP:01140-070 -- S\~{a}o Paulo, Brasil\\
}

\vskip 1cm
\end{center}

\begin{abstract}
The existence of a ``bonus'' U(1) level-one Yangian symmetry of N=4
super-Yang-Mills has recently been proposed. We provide evidence for
this proposal by constructing the BRST-invariant nonlocal charge
in the pure spinor sigma model corresponding to this bonus level-one symmetry.
We also construct analogous charges for bonus U(1) symmetries at all odd 
levels of the Yangian.
\end{abstract}

\end{titlepage}

\section{Introduction}
The four-dimensional $N=4$ supersymmetric Yang-Mills theory is invariant
under the global superconformal group $PSU(2,2|4)$. This symmetry is
manifest in the super-Yang-Mills equations of motion. Besides these global symmetries, there are also
``hidden'' non-obvious symmetries which are often referred to as Yangian
symmetries. It has been assumed that there are as many of these hidden
symmetries as there are generators in the Yangian algebra of $psu(2,2|4)$. But 
recently it was proposed in \cite{Beisert:2011pn} that there is an additional ``bonus''
Yangian symmetry at level one and  evidence for this symmetry was given in
\cite{Matsumoto:2007rh,Matsumoto:2008ww,Regelskis:2011ab}. 

In the present paper we will construct an infinite family of bonus Yangian charges 
in the corresponding pure spinor superstring sigma-model. These additional 
charges  extend the $PSU(2,2|4)$ Yangian symmetries to $PU(2,2|4)$, but only at 
the odd levels. By ``level'' we mean the number of integrals in the
most nonlocal term, minus one; for example the superconformal symmetries are
at level zero because they are given by a single integral of a local current:
\begin{equation}\label{SingleIntroduction}
\int j
\end{equation}
The bonus charges only exist at the odd levels; in particular, the global 
superconformal symmetries remain $PSU(2,2|4)$, they do not get extended to 
$PU(2,2|4)$. If this additional symmetry were a global superconformal 
symmetry like (\ref{SingleIntroduction}), then it would be the "bonus" symmetry described in \cite{Intriligator:1998ig}. 

The multilocal conserved charges were constructed in the pure spinor 
formalism in \cite{Vallilo:2003nx,Berkovits:2004xu}. For example, the following formula was used to describe
the conserved charges at level one:
\begin{equation}\label{BilocalIntroduction}
\int\int_{\sigma_1>\sigma_2} [j(\sigma_1),j(\sigma_2)] - \int k
\end{equation}
where $j$ are the currents of the global superconformal symmetry,
and $k$ is some local expression which is needed for the BRST invariance and
the conservation. Notice that $j$ takes values in the Lie superalgebra
of the superconformal symmetries, which is $psu(2,2|4)$. Therefore the
standard interpretation of the bilocal charges (\ref{BilocalIntroduction}) implies that they
take value in $psu(2,2|4)$. But nothing apriori prevents us from considering 
the commutator in (\ref{BilocalIntroduction}) as an element of $su(2,2|4)$ --- the central extension 
of $psu(2,2|4)$. Then we obtain the bilocal charges with values in $su(2,2|4)$. 
The new thing here is that we remember the central term which has 
previously been factored out. Now we have to try to repeat the 
arguments of \cite{Vallilo:2003nx,Berkovits:2004xu} and verify that they still hold even if we do keep the 
central term. We will do it in the present paper. It turns out that the arguments remain valid
for the odd level charges like (\ref{BilocalIntroduction}), but not for the even level 
charges like (\ref{SingleIntroduction}). 

There are two popular methods for packaging the higher conserved charges.
One is to use the multiple integrals like (\ref{BilocalIntroduction}), and the other is to use the 
monodromy matrix $T(z)$, which depends on the spectral parameter $z$. The 
calculations are often more automatic using $T(z)$, but explicitly analyzing 
the multiple integrals sometimes leads to interesting observations about
the BRST complex. We use both approaches in this paper. In Section \ref{sec:Monodromy} we
study the lift of $T(z)$ to the centrally extended group, and obtain the
generating function of the bonus charges in Section \ref{sec:ConstructionOfCharges}. 
Then in Section \ref{sec:ContactTerms} we study the multiple integrals and the contact terms
which arise in their BRST variation. We find that the construction of \cite{Vallilo:2003nx,Berkovits:2004xu} 
works including the central terms at the level 1, but (as expected) does not 
work at the level 0. As a biproduct we make observations about the BRST 
complex at low momentum, which may be useful for understanding the dilaton 
vertex and the $b$ ghost.

An obvious question is 
to find a physical meaning for the bonus Yangian symmetries. Although 
Yang-Mills theory at nonzero coupling is not invariant under the global 
bonus symmetry, free Yang-Mills theory is invariant since the free action is 
quadratic and preserves helicity. The sigma model corresponding to free 
Yang-Mills is at zero $AdS$ radius and was conjectured in \cite{Berkovits:2008ga} to be described
by a topological string. Turning on the vertex operator corresponding to the 
radius modulus deforms this topological sigma model and spontaneously breaks 
the $U(1)$ bonus symmetry.

One possibility is that the bonus Yangian symmetries at odd levels survive this spontaneous 
symmetry breaking. If this mechanism of spontaneous $U(1)$ symmetry breaking 
is correct, the nonlocal conserved charges found in this paper might 
be analogous to the charges found by Yoneya in \cite{Yoneya:1999qe} which describe the 
spontaneous breaking of spacetime supersymmetry. It is interesting to point 
out that the $AdS_5\times S^5$ sigma model is invariant under a discrete version of 
the bonus U(1) symmetry which switches the fermionic currents $J_1$ with $J_3$ and 
simultaneously switches the worldsheet coordinates $z$ and $\bar z$.

\paragraph     {Comment on terminology, to avoid confusion}
Expressions of the form (\ref{BilocalIntroduction}) take values in the Lie superalgebra $su(2,2|4)$.
To get a number from such an expression (rather than an element of a 
superalgebra), we must trace it with an element $\xi\in pu(2,2|4)$:
\begin{equation}
I_{\xi} =\mbox{Str}\left( \xi\;\;
(\int\int_{\sigma_1>\sigma_2} [j(\sigma_1),j(\sigma_2)] - \int k )
\;\;
\right)
\end{equation}
Similarly for the global charges:
\begin{equation}
q_{\xi} = \mbox{Str}\left( \xi\;\;
\int j
\;\;\right)
\end{equation}
We would like to stress that although (\ref{SingleIntroduction}) and (\ref{BilocalIntroduction}) {\em take values} 
in $su(2,2|4)$, the corresponding charges are {\em parametrized} by elements 
$\xi$ of $pu(2,2|4)$. The Poisson bracket of $q_{\xi}$ and $q_{\eta}$ is proportional to $q_{[\xi,\eta]}$.
The Poisson bracket of $I_{\xi}$ and $q_{\eta}$ is proportional to $I_{[\xi,\eta]}$. Therefore 
terminologically it is more natural to say that the nonlocal charges are 
extended (at the odd levels) to $PU(2,2|4)$. (Rather than to say $SU(2,2|4)$.)

\section{Notations and brief review}
\subsection{Pure spinor sigma-model}
The matter degrees of freedom are encoded in the group-valued variable 
$g\in PSU(2,2|4)$. The action is constructed out of the right-invariant 
current
\begin{equation}
\label{ric}
J =-d gg^{-1}
\end{equation}
which is invariant under $g\rightarrow gH$, with $H \in PSU(2,2|4)$ being a 
global parameter.

We use notations from Section 2 of \cite{Mikhailov:2007mr}. The  spectral parameter of 
the Lax operator will be denoted $z$. The Lax equation is
\begin{equation}
\left[ \partial_+ + J_+^{[z]}\;,\; \partial_- + J_-^{[z]} \right] = 0
\end{equation}
where
\begin{align}
J_+^{[z]}  = & J_{0+} - N_+ + z^{-1} J_{3+} + z^{-2} J_{2+} 
+ z^{-3} J_{1+} + z^{-4} N_+
\label{JPlusSpectral}
\\
J_-^{[z]}  = & J_{0-} - N_- + z J_{1-} + z^2 J_{2-} 
+ z^3 J_{3-} + z^4 N_-
\label{JMinusSpectral}
\end{align}
and $N_+$ and $N_-$ are Lorentz currents constructed out of the left and 
right-moving pure spinor ghosts. We will also introduce $l$ by
\begin{equation}
l=\log z
\end{equation}
The BRST symmetry acts on the fundamental field $g(\tau^+,\tau^-)$ in the 
following way:
\begin{equation}\label{BRSTOfG}
 Q g =  (\lambda_3 + \lambda_1) g
\end{equation}
where $\lambda_3$ and $\lambda_1$ are constructed from the left and right-moving
pure spinor ghosts.
The BRST variation of the Lax connection is given by the following expression:
\begin{equation}\label{BRSTVariationOfCapitalJ}
\epsilon QJ^{[z]} = -D^{[z]}\left(z^{-1}\epsilon \lambda_3 + z\epsilon\lambda_1\right)
\end{equation}
This is true when the equations of motion for $\lambda$ are satisfied; but the 
matter equations of motion are not needed for (\ref{BRSTVariationOfCapitalJ}).

\subsection{Nonlocal conserved charges}
\label{sec:ReviewOfConservedCharges}
Let us define the transfer-matrix in the following way:
\begin{equation}\label{DefinitionOfT}
T(z) = g(+\infty)^{-1}\; \left(P\;\exp \int \left(-J^{[z]}\right)\right) \; g(-\infty)
\end{equation}
Observe that:
\begin{equation}
-\left.{d\over dl}\right|_{l=0} T(z) = 
\int_{-\infty}^{+\infty} j
\end{equation}
where $j$ is the  density of the global conserved charges:
\begin{equation}\label{GlobalCurrent}
j= \left. g^{-1} \frac{dJ}{ dl}\right|_{l=0} g
\end{equation}
The BRST variation of the current is:
\begin{equation}
\epsilon Qj = d\left(g^{-1}\left( 
      \epsilon\lambda_3 - \epsilon\lambda_1 
\right)g\right)
\end{equation}
Similarly we may construct the higher ``Yangian-type'' currents by 
considering the higher derivatives with respect to the spectral parameter:
\begin{equation}\label{DerivativesOfT}
(-1)^n\left.{d^n\over dl^n}\right|_{l=0} T(z)
= \int j \ldots\int j
\end{equation}
The right hand side is the sum of multiple integrals; the leading term is the 
$n$-tuple integral of $n$ global currents (\ref{GlobalCurrent}), and the subleading terms are 
the integrals of lower multiplicity. We observe:
\begin{align}
&\epsilon Q {d^n\over dl^n} T(z) = 
\nonumber
\\   
= & \;{d^n \over dl^n}\left(
   g^{-1}( +\infty)\; P\left[
      \left(\int_{-\infty}^{+\infty} 
          D^{[z]}\left(\epsilon\lambda^{[z]}\right)\right)
      \exp \int_{-\infty}^{+\infty} 
         \left(-J^{[z]}\right)
      \right]\; g(-\infty)
\right) =
\nonumber \\  
= & \; \mbox{\tt boundary terms}
\end{align}
so it is automatically BRST-closed.

\section{From $PSU(2,2|4)$ to $SU(2,2|4)$}\label{sec:Monodromy}
The worldsheet $g$ takes values in the group manifold $PSU(2,2|4)$.
Therefore the transfer matrix $T$ also takes values in $PSU(2,2|4)$.

\subsection{Lifting $g$}
Let us, however, consider some arbitrary lift of $g$ from $PSU(2,2|4)$ 
to $SU(2,2|4)$. After we lift $g$, it makes sense to consider  $T$ 
defined in (\ref{DefinitionOfT}) also as an element of $SU(2,2|4)$.
It is natural to ask the following questions:
\begin{enumerate}
\item there are many ways to lift $g$ from $PSU$ to $SU$; is $T$ independent of the 
choice of the lift of $g$?
\item is it true that the lifted $T$ is BRST-closed?
\item is it true that the lifted $T$ does not depend on the contour?
\end{enumerate}
It turns out that the answers to all of these questions are negative.
Let us for example consider the first question:
\begin{itemize}
\item what happens when we change $g \to e^{i\phi} g$ where $\phi$ is a real-valued 
function of $\tau^+$ and $\tau^-$?
\end{itemize}
It turns out that $T$ {\em does} change. The variation comes from the
variation of $J_2 = -(dg g^{-1})_{\bar{2}}$, and equals to the following expression:
\begin{equation}\label{GaugeVariationOfT}
T(z) \to \exp\left[ {i\over 2} \left( 
   \left(
      z^{-2} - z^2
   \right) \int *d\phi  
+  \left(
      z^{-2} + z^2
   \right) \int d\phi  
\right) \right]\;T(z)
\end{equation}
The second term in the parenthesis is reduced to boundary terms,
but the first one is an essential bulk term. 

Notice that most of the components of $T(z)$ do not depend on the choice of a 
lift, only the ``central component'' does. To capture this 
``central component'', we will introduce the following notation:
\begin{align}
C(z)\;  \stackrel{\mbox{\tiny \tt def}}{=} \;&\;
\mbox{Str}\left(
      s\log T(z)
\right)
\\[7pt]
\mbox{\small \tt where } s =\; & \left(
   \begin{array}{cc}
   {\bf 1}_{4\times 4} & 0 \cr
   0 & -{\bf 1}_{4\times 4}
\end{array}
\right)
 \nonumber
\end{align}
Let us think of $T(z)$ as an $4|4 \times 4|4$ matrix. We understand $\log T(z)$ as a 
power series expansion in $l$ around $l=0$; when $l=0$ we get $T(z=1)={\bf 1}$. 

Equation (\ref{GaugeVariationOfT}) implies that $C(z)$ is not a well-defined function on the 
string phase space, because it does depend on the lift of $g$. But the
dependence on the choice of the lift is rather simple. When we change
$g \to e^{i\phi} g$ we get:
\begin{align}
\label{ChangeOfCOfZ}
C(z) \to\;\; & C(z) 
+ 4i \left( 
   \left(
      z^{-2} - z^2
   \right) \int *d\phi  
+  \left(
      z^{-2} + z^2
   \right) \int d\phi  
\right) 
\end{align}
The second term in parentheses is a total derivative, but the first one
is not.

\subsection{Construction of additional charges}
\label{sec:ConstructionOfCharges}
Let us pick $N$ complex numbers $z_1,\ldots,z_N$ satisfying:
\begin{equation}
\sum\limits_{i=1}^N(z^{-2}_i - z^2_i)  = 0
\end{equation}
and consider $I(z_1,\ldots,z_N)$ defined as follows:
\begin{equation}\label{IOfZs}
I(z_1,\ldots,z_N) = \sum\limits_{i=1}^N \mbox{Str} \; C(z_i) 
= \sum\limits_{i=1}^N \mbox{Str}
\left(s\;\log T(z_i)\right)
\end{equation}
Eq. (\ref{ChangeOfCOfZ}) implies that up to boundary terms $I(z_1,\ldots,z_N)$ does not depend on 
the choice of  a lift. In particular, let us consider $N=2$, $z_1=z$ and 
$z_2=z^{-1}$. Define $I(l)$ as follows:
\begin{align}
I(l)
&
=\mbox{Str}\left( \, s\left[ 
      \; \log T(z) + \log T(z^{-1})\, 
\right]\;\right)
\end{align}
Expanding in powers of $l$ we get:
\begin{align}
I(l) = &\; I_2 l^2 + I_4 l^4 + I_6 l^6 + \ldots \; 
\\[7pt]  
\mbox{\small \tt where }\;
I_2  = &\; \mbox{Str}\left(s \left[T'' -  (T')^2\right]\right)
\label{I2} \\   
I_4  = &\; \mbox{Str}\left(s \left[
      {1\over 12}T'''' - {1\over 6} (T'T''' + T'''T') - {1\over 4}(T'')^2 +
\right.\right.
\nonumber \\    
&\phantom{Str aaa}
\left.\left. + {1\over 3}\left((T')^2T'' + T'T''T' + T''(T')^2\right)
  - {1\over 2}(T')^4
\right]\right)
\\  
&\mbox{\small \tt etc. }\; 
\end{align}
where $T' = \left.{dT\over dl}\right|_{l=0}$,\hspace{10pt} $T'' = \left.{d^2 T\over dl^2}\right|_{l=0}$, \hspace{5pt} {\it etc.}
This gives an infinite family of bonus charges.

\subsection{Independence of the deformation of the contour}
The equation of motion for currents is only satisfied modulo the central 
terms. After the lift we have:
\begin{equation}
[D_+^{[z]}\;,\;D_-^{[z]}]  = \left( {1\over z^2} - z^2 \right) F_{+-}
\end{equation}
where $F = F_{+-}\;d\tau^+\wedge d\tau^-$ is some $z$-independent 2-form taking values in the
center of $su(2,2|4)$. This implies that the variation of $T$ under the change 
of the contour $C$ is:
\begin{equation}
\delta T(z) = \left[-\int_C \left( {1\over z^2} - z^2 \right)
\iota_{v} F \right] T(z)
\end{equation}
where $v$ is the section of the normal bundle to the contour, which describes 
its variations. This implies that the variation of (\ref{IOfZs}) is zero, {\it i.e.}
that (\ref{IOfZs}) is contour-independent.

\section{BRST symmetry and contact terms}\label{sec:ContactTerms}
In this Section we will discuss the construction of the bonus charges 
from the point of view of \cite{Berkovits:2004xu}.

\subsection{Nonlocal charges from multiple integrals}
\subsubsection{Definition and transformation under conformal supersymmetries}
Let us try to construct the bonus charge as in \cite{Berkovits:2004xu}:
\begin{align}\label{SumOfDoubleAndSingle}
\mbox{Str}\left(
s \int\int_{\sigma_1<\sigma_2} [j(\sigma_1)\;,\;j(\sigma_2)]
\right) - \int (\ldots)
\end{align}
where: 
\begin{equation}
s = \left(\begin{array}{cc} {\bf 1}_{4\times 4} & 0 \cr 
0 & - {\bf 1}_{4\times 4} \end{array}\right)
\end{equation}
and $j(\sigma)$ are the conserved currents for the global $PSU(2,2|4)$ symmetries.
Contact terms prevent the double integral in the first term of (\ref{SumOfDoubleAndSingle}) from 
being contour-independent on its own. But sometimes it is possible to 
compensate its variation by the variation of the second term, which is a 
single integral of a local expression. 

Under global $PSU(2,2|4)$ transformations parameterized by $\Omega$, the first 
term of  (\ref{SumOfDoubleAndSingle}) transforms into
\begin{align}\label{SumtwoOfDoubleAndSingle}
\mbox{Str}\left(
s ~[\Omega, ~\int\int_{\sigma_1<\sigma_2} [j(\sigma_1)\;,\;j(\sigma_2)]
~] \right) = 
\mbox{Str}\left(
[s,\Omega] ~\int\int_{\sigma_1<\sigma_2} [j(\sigma_1)\;,\;j(\sigma_2)]
 \right) 
\end{align}
where $[s,\Omega] =\pm 2\;\Omega$ if $\Omega$ is a fermionic $PSU(2,2|4)$ generator and 
$[s,\Omega]=0$ if $\Omega$ is a bosonic $PSU(2,2|4)$. So (\ref{SumOfDoubleAndSingle}) transforms as the 
level-one bonus Yangian generator as described in \cite{Beisert:2011pn}.

\subsubsection{BRST variation}
Since $\int j(\sigma)$ is BRST-closed, $\epsilon Q~j = d~\Lambda(\epsilon)$ for some $\Lambda(\epsilon)$ 
satisfying $Q\Lambda(\epsilon) = 0$. Using the notation of \cite{Berkovits:2004jw}, one finds that
\begin{equation}\label{DefLambda}
\Lambda(\epsilon) = g^{-1} (\epsilon\lambda_3 - \epsilon\lambda_1) g 
\end{equation}
The BRST variation of the double integral therefore gives:
\begin{equation}\label{SingleIntegral}
2\;\mbox{Str}\left(
   s\int \left[\Lambda(\epsilon), j\right]
\right)
\end{equation}
The double integral became a single integral. As a check, let us verify 
that (\ref{SingleIntegral}) is $Q$-exact:
\begin{align}
2\;\epsilon Q\;\mbox{Str}\left(
   s\int \left[\Lambda(\epsilon'), j\right]
\right) =& \;
2\;\mbox{Str}\left(
    s\int \left[\Lambda(\epsilon'), d\Lambda(\epsilon)\right]
\right) =
\nonumber \\   
=& \;
d\;\mbox{Str}\left(
    s\int \left[\Lambda(\epsilon'), \Lambda(\epsilon)\right]
\right) = 
\nonumber \\  
=& \;\mbox{\small \tt boundary terms}
\end{align}
As was explained in \cite{Berkovits:2004xu}, the construction of multilocal charges requires that 
the following expression:
\begin{equation}\label{follow}
\mbox{Str}\left(
   s\;\left[\Lambda(\epsilon'), \Lambda(\epsilon)\right]
\right)
\end{equation}
is $Q$-exact. Using (\ref{DefLambda}),
\begin{align}
\mbox{Str}\left(
   s\;\left[\Lambda(\epsilon'), \Lambda(\epsilon)\right]
\right) = 
2\epsilon'\epsilon\;
\mbox{Str}\left(
 s~g^{-1}~(\lambda_3 \lambda_3 + \lambda_1\lambda_1 - \{\lambda_1,\lambda_3\}) g
\right) =\nonumber \\      
= -  \;\epsilon Q\; 
\mbox{Str}\left(
 s~g^{-1}~(\epsilon'\lambda_1+ \epsilon'\lambda_3)~g
\right) + 
4\epsilon'\epsilon~\mbox{Str}\left(
 s~g^{-1}~(\lambda_1^2 + \lambda_3^2)~g
\right)
\end{align}
The pure spinor constraint implies that $\lambda_1^2$ and $\lambda_3^2$ are in the center 
of $su(2,2|4)$. So (\ref{follow}) is equal to the sum of a $Q$-exact expression:
\begin{equation}
- \epsilon Q\;
\mbox{Str}(s~g^{-1}(\epsilon'\lambda_3 + \epsilon'\lambda_1)g)
\end{equation}
{\bf plus} the following expression, which does not contain $g$:
\begin{equation}\label{CenterObstacle}
4\epsilon'\epsilon \;
\mbox{Str}\left(s\lambda_3^2 + s\lambda_1^2\right) 
\end{equation}
It will now be shown that (\ref{CenterObstacle}) is $Q$-exact.

\subsubsection{BRST triviality of the central terms}
Equation (\ref{CenterObstacle}) can be rewritten either using the notations of \cite{Mikhailov:2011af}:
\begin{equation}\label{InMyNotations}
\mbox{Str}\left(\lambda_3[s,\lambda_3] + \lambda_1[s,\lambda_1]\right) =
-2i\left(||\lambda_L\cap\lambda_L|| - ||\lambda_R\cap\lambda_R||\right)
\end{equation}
or the gamma-matrix notations:
\begin{equation}
(\lambda_L^\alpha\widehat{F}_{\alpha\beta}\lambda_L^\beta) -
(\lambda_R^\alpha\widehat{F}_{\alpha\beta}\lambda_R^\beta)
\end{equation}
where $\alpha=1$ to 16 is an SO(9,1) spinor index, $\widehat{F}_{\alpha\beta}$ is the RR 5-form field 
strength contracted with five $\Gamma$-matrices, and $\lambda_L^\alpha$ and $\lambda_R^\alpha$ are the left and 
right-moving $d=10$ pure spinors. The relation to the notations of \cite{Mikhailov:2011af} is:
\begin{align}
\lambda_3 =\left(\begin{array}{cc}
0 & \lambda_L \cr
i \omega^{-1} \lambda_L\omega & 0
\end{array}\right)
\quad\mbox{ \small \tt and }\quad 
\lambda_1 =\left(\begin{array}{cc}
0 & \lambda_R \cr
- i \omega^{-1} \lambda_R\omega & 0
\end{array}\right)
\end{align}
The expression (\ref{CenterObstacle}) is obviously BRST-closed, but it is also BRST-exact. 
In order to demonstrate that it is BRST exact, we consider a special class of 
ghost number 1 vertices which was introduced in \cite{Berkovits:2007rj}.

Consider the following decomposition of the Lie superalgebra ${\bf g} = u(2,2|4)$:
\begin{equation}\label{TriangularDecompositionOfAlgebra}
{\bf g} = {\bf g}_+ + {\bf g}_{\mbox{\small\tt even}} + {\bf g}_-
\end{equation}
where ${\bf g}_{\mbox{\small\tt even}}$ is the even subalgebra and ${\bf g}_+$ and ${\bf g}_-$ consist of the matrices 
of the form $\left(\begin{array}{cc} 
0_{4\times 4} & X_{4\times 4} \cr 0_{4\times 4} & 0_{4\times 4} 
\end{array}\right)$ and $\left(\begin{array}{cc} 
0_{4\times 4} & 0_{4\times 4} \cr X_{4\times 4} & 0_{4\times 4}  
\end{array}\right)$, respectively.
We start with a variant of the Gauss decomposition of $g$:
\begin{align}\label{GaussDecomposition}
&g  = \; e^{\theta_+} e^x e^{\theta_-}
\\    
& \mbox{\small \tt where}\;\;
\theta_{\pm}\in {\bf g}_{\pm}\;,\;\;
x\in {\bf g}_{\rm even}
\end{align}
We then define the ghost number one expression as follows:
\begin{equation}\label{VGLambda}
V(g,\lambda,\epsilon) = 
2\;\mbox{Str}\left(
  \; s\;[\;\epsilon \lambda_3 + \epsilon\lambda_1\;,\;\theta_+]\;
\right)
\end{equation}
In the gamma-matrix notation this equals to:
\begin{equation}
V(g,\lambda) = 
(\lambda_L\widehat{F}\theta_+) - (\lambda_R\widehat{F}\theta_+)
\end{equation}
Then $Q$ of this $V(g,\lambda,\epsilon)$ equals (\ref{CenterObstacle}):
\begin{equation}\label{QVIsLambdaLambda}
\epsilon' Q V(g,\lambda,\epsilon) = 2\epsilon\epsilon' \;
\mbox{Str}\left(s\lambda_3^2 + s\lambda_1^2\right) 
\end{equation}
It is not immediately obvious why the RHS of (\ref{QVIsLambdaLambda}) does not contain terms 
quadratic or higher power in $\theta$. We will prove it using the results of \cite{Mikhailov:2011af}.
\anth{Lemma:}
\begin{equation}\label{QLambdaThetaMinusMuTheta}
Q(||\lambda_L\cap\theta_+|| - ||\lambda_R\cap\theta_+||) =
||\lambda_L\cap\lambda_L|| - ||\lambda_R\cap\lambda_R||
\end{equation}
\anth{Proof:} 
The action of BRST symmetry in the coordinates (\ref{GaussDecomposition}) was computed in \cite{Mikhailov:2011af}:
\begin{align}
\epsilon Q_{L} \Phi = &\; 
\left(\epsilon\lambda_L{\partial\over\partial\theta_+}\right) \Phi 
- i \left(
(\theta_+\cap\epsilon\lambda_L\cup\theta_+){\partial\over\partial\theta_+}
\right)\Phi
\label{QLPhi}
\\   
\epsilon Q_{R} \Phi = &\; 
\left(\epsilon\lambda_R{\partial\over\partial\theta_+}\right) \Phi 
+ i \left(
(\theta_+\cap\epsilon\lambda_R\cup\theta_+){\partial\over\partial\theta_+}
\right)\Phi
\label{QRPhi}
\end{align}
Then (\ref{QLambdaThetaMinusMuTheta}) follows from the following observations:
\begin{align}
||\lambda_L\cap\theta_+\cup\lambda_R\cap\theta_+|| = & \; 0
\\  
||\lambda_L\cap\theta_+\cup\lambda_L\cap\theta_+|| + 
||\lambda_R\cap\theta_+\cup\lambda_L\cap\theta_+|| = & \; 0
\\   
||\lambda_R\cap\theta_+\cup\lambda_R\cap\theta_+|| = & \; 0
\end{align}
which follow from symmetries. In  the gamma-matrix notations, the vanishing 
of these terms can be seen in the following way. Observe that the second 
term on the RHS of (\ref{QLPhi}) can be written as:
\begin{equation}
Q_{L} \theta_+  = -i \widehat{F} \Gamma^m\theta_+ (\theta_+\Gamma_m\lambda_L)
\end{equation}
Similarly, $Q_{R}\theta_+$ results in a similar expression but with the overall minus
sign. Then we get, {\it e.g.}, 
\begin{equation}
(Q_{L}+Q_R)((\lambda_L-\lambda_R)\widehat{F}\theta_+) 
= -i (\theta_+ \Gamma_m(\lambda_L-\lambda_R)) 
((\lambda_L-\lambda_R) \widehat{F}^2 \Gamma^m \theta_+) 
\end{equation}
\begin{equation}
= -i (\theta_+ \Gamma_m(\lambda_L-\lambda_R))(\theta_+ \Gamma^m
(\lambda_L-\lambda_R))  = 0
\end{equation}
(This would be zero even if $\lambda$ were not pure, just because of the symmetry.)

This means that the following expression:
\begin{align}
2\;\mbox{Str}\left(s\; 
      [\Lambda(\epsilon) , j]  
\right)
      +\; d~\mbox{Str}
\left(s g^{-1}(\epsilon \lambda_3 + \epsilon \lambda_1)g\right)
 -
\nonumber \\    
- \; d\left(
   (\epsilon\lambda_L\widehat{F}\theta_+) 
   - (\epsilon\lambda_R\widehat{F}\theta_+)
\right)
\label{QClosedDimensionOne}
\end{align}
is $Q$-closed, and therefore also $Q$-exact since the BRST cohomology is 
trivial for operators of non-zero conformal weight. 
The expression (\ref{QClosedDimensionOne}) can be also presented in the following way:
\begin{align}
2\;\mbox{Str}\left(s\; 
      [\Lambda(\epsilon) , j]  \right)
      + \; d\;\mbox{Str}
\left(s g^{-1}(\epsilon \lambda_3 + \epsilon \lambda_1)g\right)
-
\nonumber \\    
- \;4\; d \; \mbox{Str}\; \left(
s [\epsilon\lambda_3, \theta_+] + s [\epsilon\lambda_1,\theta_+]
\right)
\label{QClosedDimensionOneAlt}
\end{align}
Setting (\ref{QClosedDimensionOne}) equal to $Q(...)$, one finds that (\ref{SumOfDoubleAndSingle}) is BRST-closed.

The last term in (\ref{QClosedDimensionOne}) appears rather mysterious. 
It cannot be expressed in manifestly $PSU(2,2|4)$ covariant notation, so it is unclear
if the arguments
of \cite{Berkovits:2004xu} can be used to prove that the bonus Yangian symmetries
survive after including quantum corrections.
Furthermore, the operator 
\begin{equation}\label{dilaton}
(\lambda_L \widehat{F}\lambda_L - \lambda_R\widehat{F}\lambda_R)=
(Q_L+Q_R)
  \left( (\lambda_L\widehat{F}\theta_+) - (\lambda_R\widehat{F}\theta_+) \right)
\end{equation}
bears a certain resemblence to the zero-momentum dilaton vertex operator
in bosonic string theory which can be expressed as \cite{Bergman:1994qq}
\begin{equation}\label{Vdil}
V_D = (Q_L+Q_R) (\partial c_L - \bar\partial c_R) = c_L \partial^2 c_L
- c_R \partial^2 c_R
\end{equation}
where $c_L$ and $c_R$ are the left and right-moving reparameterization
ghosts. Note that $V_D$ is not BRST-trivial in the semirelative cohomology,
i.e. if one requires that the zero model of $b_L-b_R$
annihilates the gauge parameter. It would be interesting
to study further the resemblence of (\ref{dilaton}) and (\ref{Vdil})
using the $AdS_5\times S^5$ $b$ ghost constructed in \cite{Berkovits:2010zz}.
However, in this paper, we will explain the structure of (\ref{dilaton})
in a different way in the following Section \ref{sec:LiftedBRST}.

\subsection{Lifted BRST operator and compensating gauge transformation}
\label{sec:LiftedBRST}
\subsubsection{BRST transformation of the components of currents}
With the {\em lifted} $g$ and the BRST operator still defined by
(\ref{BRSTOfG}), we have:
\begin{align}
\epsilon Q \left( g^{-1} \frac{d^2J^{[z]}}{ dl^2} g\right) 
 = & \;
  g^{-1} \left[ 
     \frac{d^2 J^{[z]}}{ dl^2},\; \epsilon \lambda^{[z]} 
  \right] g
- g^{-1} \frac{d^2}{ dl^2} (D^{[z]}\epsilon\lambda^{[z]}) g =
\nonumber 
\\
 = & 
-2g^{-1} \left[ 
   \frac{dJ^{[z]}}{ dl},\; \frac{d\epsilon\lambda^{[z]}}{ dl} 
\right] g
-d(g^{-1} \epsilon\lambda g)
\label{DescentOfCommutatorMiddleStep}
\end{align}
where the left hand side and the right hand side are both taken at $z=1$.

Equation (\ref{DescentOfCommutatorMiddleStep}) is the same when $g$ and $J$ are lifted to $SU(2,2|4)$ and
$su(2,2|4)$, as it was when we considered them as elements of $PSU(2,2|4)$ 
and $psu(2,2|4)$. What changes, however, is that after the lift the BRST 
operator defined by (\ref{BRSTOfG}) is not nilpotent:
\begin{equation}
Q^2 g = \left(\lambda_3^2 + \lambda_1^2\right) g + \{\lambda_3,\lambda_1\} g
\end{equation}
This results in the following variation of $J^{[z]}$:
\begin{equation}
Q^2 J^{[z]} = - d\left( z^{-2}\lambda_3^2 + z^2\lambda_1^2\right) 
- D^{[z]}\{\lambda_3,\lambda_1\}
\end{equation}
This means that the right hand side of (\ref{DescentOfCommutatorMiddleStep}) is not BRST-closed\footnote{The 
last term $-d\{\lambda_3,\lambda_1\}$ was present even before the lift, but
it only results in the boundary terms. 
The new terms $-d\left( z^{-2}\lambda_3^2 + z^2\lambda_1^2\right)$ are $z$-dependent; this is because the unit matrix has
${\bf Z}_4$-grading $\bar{2}$, rather than $\bar{0}$}.

\subsubsection{Fixing the gauge}
The lifted $Q$ is not nilpotent, but its square is the $u(1)$ gauge
transformation. Therefore we can make the lifted $Q$ nilpotent if we fix the 
$u(1)$ gauge, {\it i.e.} choose a particular lift.
The following gauge is convenient:
\begin{align}\label{GaugeCondition}
&g  = \; e^{\theta_+} e^x e^{\theta_-}
\\    
& \mbox{\small \tt where}\;\;
\theta_{\pm}\in {\bf g}_{\pm}\;,\;\;
x\in {\bf g}_{\rm even}\;,\;\; \mbox{Str}\;(sx) = 0
\end{align}
where the notations are the same as in (\ref{TriangularDecompositionOfAlgebra}). What actually fixes the gauge
is the condition $\mbox{Str}\;(sx) = 0$.

We modify $Q$ by adding to it the compensating gauge transformation, so that 
the resulting $\widehat{Q}$ preserves the gauge: 
\begin{equation}
\widehat{Q} g = \lambda g + c(g,\lambda) g
\end{equation}
where $c(g,\lambda)$ is an appropriate $u(1)$ gauge compensator, proportional to
the unit matrix.  This compensated $\widehat{Q}$ is automatically nilponent:
\begin{equation}
\widehat{Q}^2 = 0
\end{equation}
which implies that the BRST variation of $c$ is given by:
\begin{equation}
Qc = - (\lambda_3^2 + \lambda_1^2)
\end{equation}
We observe that the variation of $J^{[z]}$ under the compensating gauge 
transformation is given by:
\begin{equation}\label{CompensatingGaugeTransformation}
\delta_{\rm comp}J^{[z]} = - {1\over 2}\left(
   (z^{-2} + z^2) dc + (z^{-2} - z^2) *dc
\right)
\end{equation}
\subsubsection{Lifted global currents are not BRST closed}
\label{sec:LiftingGlobal}
Let us try to lift the conserved current corresponding to the
global symmetry, using the gauge (\ref{GaugeCondition}).
We observe that the BRST variation of the lifted global conserved current 
is not a total derivative:
\begin{align}
\widehat{Q} j = 
\widehat{Q} \left( g^{-1}{dJ\over dl} g\right) = & 
\mbox{ [total derivative] } + 2*dc
\end{align}
The extra term $2 * dc$ comes from the compensating gauge transformation
(\ref{CompensatingGaugeTransformation}) of $J_{\bar{2}}$.
We conclude that we cannot lift the local conserved charges corresponding
to the global symmetries. Therefore the global symmetries are indeed only 
$PSU(2,2|4)$, and not $SU(2,2|4)$. 

Notice also that in the {\it bilocal} charges (\ref{SumOfDoubleAndSingle}), in the variation of 
the double integral, the term $2 * dc$ enters as the commutator $[j\;,\; 2*dc]$, 
and this vanishes because $c$ is in the center. 

\subsubsection{Explanation of Eq. (\ref{QClosedDimensionOneAlt})}
Let us now return to the discussion of the contact term in the BRST variation
of the bilocal charges. We have the following modification of (\ref{DescentOfCommutatorMiddleStep}):
\begin{align}
\epsilon \widehat{Q} \left( g^{-1} \frac{d^2J^{[z]}}{ dl^2} g\right) 
 =  & 
-2g^{-1} \left[ 
   \frac{dJ^{[z]}}{ dl},\; \frac{d\epsilon\lambda^{[z]}}{ dl} 
\right] g
-d(g^{-1} \epsilon\lambda g) -
\nonumber
\\  
& - 4dc
\label{DescentOfCommutatorMiddleStepLifted}
\end{align}
Let us now study the compensator $c$ more explicitly. A calculation using the 
gauge (\ref{GaugeCondition}) gives:
\begin{equation}
c(g,\epsilon\lambda) = - \frac{
  \mbox{Str}\left(s[\epsilon\lambda_3,\theta_+]\right)
  + \mbox{Str}\left(s[\epsilon\lambda_1,\theta_+]\right)
}{ \mbox{Str}\; s }
\end{equation}
This means that (\ref{DescentOfCommutatorMiddleStepLifted}) agrees with (\ref{QClosedDimensionOneAlt}). This explains the last term in 
Eqs. (\ref{QClosedDimensionOne}), (\ref{QClosedDimensionOneAlt}) and shows that (\ref{QClosedDimensionOne}) equals to $\epsilon Q \left( g^{-1} \frac{d^2J^{[z]}}{ dl^2} g\right) $ 
in the gauge (\ref{GaugeCondition}).

\subsection{Summary of the double integral construction}
We conclude that the following bilocal charge is BRST-closed up to 
boundary terms, {\em including the $u(1)$ part}:
\begin{equation}\label{SummaryOnDoubleIntegral}
\int\int_{\sigma_1>\sigma_2} [j(\sigma_1),j(\sigma_2)] - \int k
\end{equation}
where $j = g^{-1}\left.{dJ\over dl}\right|_{l=0} g$ and $k=g^{-1}\left.{d^2J\over dl^2}\right|_{l=0} g$. Notice that a different choice
of a lift of $g$ would change $k$ by adding to it a total derivative, and $c$ of
(\ref{DescentOfCommutatorMiddleStepLifted}) by a BRST exact expression.

In fact (\ref{SummaryOnDoubleIntegral}) is equal to $I_2$ given by (\ref{I2}). 
The subtraction of $(T')^2$ turns $2\int\int_{\sigma_1>\sigma_2} j(\sigma_1)j(\sigma_2)$ into
$\int\int_{\sigma_1>\sigma_2}[j(\sigma_1),j(\sigma_2)]$, which is necessary for the $u(1)$ gauge invariance
(see the end of Section \ref{sec:LiftingGlobal}).

Eq. (\ref{SummaryOnDoubleIntegral}) is essentially the same as in \cite{Berkovits:2004xu}, except for in \cite{Berkovits:2004xu} it was only claimed up to the central terms, in other words up to $u(1)$. 
The current paper shows that actually the $u(1)$ part of the same 
expression (\ref{SummaryOnDoubleIntegral}) is also  BRST closed and contour-independent (up to
boundary terms).
This corresponds to the bonus symmetry of \cite{Beisert:2011pn}.

\section*{Acknowledgments}
NB would like to thank Niklas~Beisert and Juan~Maldacena for useful discussions
and CNPq grant 300256/94-9 and FAPESP grant 09/50639-2 for partial
financial support.
The work of A.M. was supported in part by 
the Ministry of Education and Science of the Russian Federation
under contract 14.740.11.0347, and in part by the Russian Foundation for
Basic Research grant RFBR10-02-01315.


\begin{thebibliography}{10}

\bibitem{Beisert:2011pn}
N.~Beisert and B.~U.~W. Schwab, {\it {Bonus Yangian Symmetry for the Planar
  S-Matrix of N=4 Super Yang-Mills}},
  \href{http://xxx.lanl.gov/abs/arXiv/1103.0646}{{\tt arXiv/1103.0646}}.

 
\bibitem{Matsumoto:2007rh}
  T.~Matsumoto, S.~Moriyama and A.~Torrielli,
  {\it ``A Secret Symmetry of the AdS/CFT S-matrix,''}
  JHEP {\bf 0709}, 099 (2007)
  [arXiv:0708.1285 [hep-th]].

\bibitem{Matsumoto:2008ww}
  T.~Matsumoto and S.~Moriyama,
  {\it ``An Exceptional Algebraic Origin of the AdS/CFT Yangian Symmetry,''}
  JHEP {\bf 0804}, 022 (2008)
  [arXiv:0803.1212 [hep-th]].

\bibitem{Regelskis:2011ab}
V.~Regelskis, {\it {The secret symmetries of the AdS/CFT reflection matrices}},
  \href{http://xxx.lanl.gov/abs/arXiv/1105.4497}{{\tt arXiv/1105.4497}}.
  
\bibitem{Intriligator:1998ig}
K.~A. Intriligator, {\it {Bonus symmetries of N = 4 super-Yang-Mills
  correlation functions via AdS duality}},  {\em Nucl. Phys.} {\bf B551} (1999)
  575--600, [\href{http://xxx.lanl.gov/abs/arXiv/hep-th/9811047}{{\tt
  arXiv/hep-th/9811047}}].

\bibitem{Vallilo:2003nx}
B.~C. Vallilo, {\it Flat currents in the classical {AdS(5) x S(5)} pure spinor
  superstring},  {\em JHEP} {\bf 03} (2004) 037,
  [\href{http://xxx.lanl.gov/abs/hep-th/0307018}{{\tt hep-th/0307018}}].

\bibitem{Berkovits:2004xu}
N.~Berkovits, {\it Quantum consistency of the superstring in {AdS(5) x S(5)}
  background},  {\em JHEP} {\bf 03} (2005) 041,
  [\href{http://xxx.lanl.gov/abs/hep-th/0411170}{{\tt hep-th/0411170}}].

\bibitem{Berkovits:2008ga}
N.~Berkovits, {\it {Simplifying and Extending the {AdS(5) x S(5)} Pure Spinor
  Formalism}},  {\em JHEP} {\bf 09} (2009) 051, 
  \href{http://xxx.lanl.gov/abs/arXiv/0812.5074}{{\tt
  arXiv/0812.5074}}.

\bibitem{Yoneya:1999qe}
T.~Yoneya, {\it {Spontaneously broken space-time supersymmetry in open string
  theory without GSO projection}},  {\em Nucl. Phys.} {\bf B576} (2000)
  219--240, [\href{http://xxx.lanl.gov/abs/arXiv/hep-th/9912255}{{\tt
  arXiv/hep-th/9912255}}].

\bibitem{Mikhailov:2007mr}
A.~Mikhailov and S.~Schafer-Nameki, {\it Perturbative study of the transfer
  matrix on the string worldsheet in {AdS(5) x S(5)}},
  \href{http://xxx.lanl.gov/abs/arXiv:0706.1525 [hep-th]}{{\tt arXiv/0706.1525
  [hep-th]}}.

\bibitem{Berkovits:2004jw}
N.~Berkovits, {\it {BRST} cohomology and nonlocal conserved charges},  {\em
  JHEP} {\bf 02} (2005) 060,
  [\href{http://xxx.lanl.gov/abs/hep-th/0409159}{{\tt hep-th/0409159}}].

\bibitem{Mikhailov:2011af}
A.~Mikhailov, {\it {Finite dimensional vertex}},
  \href{http://xxx.lanl.gov/abs/arXiv/1105.2231}{{\tt arXiv/1105.2231}}.

\bibitem{Berkovits:2007rj}
N.~Berkovits and C.~Vafa, {\it {Towards a Worldsheet Derivation of the
  Maldacena Conjecture}},  {\em JHEP} {\bf 03} (2008) 031,
  [\href{http://xxx.lanl.gov/abs/arXiv/0711.1799}{{\tt arXiv/0711.1799}}].

\bibitem{Bergman:1994qq}
O.~Bergman and B.~Zwiebach, {\it {The Dilaton theorem and closed string
  backgrounds}},  {\em Nucl. Phys.} {\bf B441} (1995) 76--118,
  [\href{http://xxx.lanl.gov/abs/arXiv/hep-th/9411047}{{\tt
  arXiv/hep-th/9411047}}].

\bibitem{Berkovits:2010zz}
N.~Berkovits and L.~Mazzucato, {\it {Taming the b antighost with Ramond-Ramond
  flux}},  {\em JHEP} {\bf 11} (2010) 019,
  [\href{http://xxx.lanl.gov/abs/arXiv/1004.5140}{{\tt arXiv/1004.5140}}].

\end{thebibliography}

\def\cprime{$'$} \def\cprime{$'$}
\providecommand{\href}[2]{#2}\begingroup\raggedright\endgroup

\end{document}